\documentstyle[12pt]{article}

\newcommand {\bean} {\begin{eqnarray*} }
\newcommand {\ean} {\end{eqnarray*} }
\newcommand {\bea} {\begin{eqnarray} }
\newcommand {\eea} {\end{eqnarray} }

\newcommand {\rfa}[1] { (\ref{#1}) }

\newcommand {\be} {\begin{equation} }
\newcommand {\ee} {\end{equation} }

\newcommand{\C}{{\sf C\hspace*{-0.9ex}\rule{0.15ex}%
{1.3ex}\hspace*{0.9ex}}}

\newcommand{\ra}{\rightarrow}

\newcommand{\lm}{\lambda}
\begin{document}

\noindent {\large\bf Interaction energy of Chern-Simons vortices
in the gauged O(3) sigma model.}
\vspace{1cm}

\noindent K. Arthur \footnote{e-mail: 93701063@tolka.dcu.ie}.
\\ School of Mathematical Sciences, Dublin City University,
\\ Dublin 9, Ireland.

\vspace{2cm}

\begin{abstract}

The purpose of this Letter is to present a computation of the interaction
energy of gauged O(3) Chern-Simons vortices which are infinitely separated.
The results will show the behaviour of the interaction energy as a function
of the constant coupling the potential, which measures the relative strength
of the matter self-coupling and the electromagnetic coupling. We find that
vortices attract each other for $\lambda > 1 $ and repel when $\lambda < 1 $.
When $\lambda =1 $ there is a topological lower bound on the energy.
It is possible to saturate the bound if the fields satisfy a
set of first order partial differential equations.
\end{abstract}

\newpage

In recent years the study of vortices has become a subject dealing with
both particle physics and condensed matter physics. The physics of the
recently discovered high-$T_{c}$ superconductors is an important problem.
The studies of Chern-Simons (C-S) solitons can be related to the unusual
behaviour of this new type of superconductor.

The purpose of this Letter is to present a computation of the interaction
energy of gauged O(3) Chern-Simons vortices.
We discuss the solutions analytically and numerically, and present the
numerical results. The corresponding problem for vortices of the Abelian Higgs
model was first considered by Jacobs and Rebbi \cite{Jacobs},
as expected, our results are analagous to these. The corresponding problem
for Abelian Chern-Simons Higgs vortices was considered in \cite{Arthur1}.

There is an example in which this crossover, from attractive to repulsive
behaviour occurs, namely, the case of Abelian Skyrme $CP^{1}$ vortices
\cite{Tchrakian}. In this case no Bogomol'nyi bound occurs,
yet crossover takes place.

The essential feature of C-S solitons of all $2+1$ dimensional theories is
that the time component of the Euler-Lagrange equation arising from the
variation of the gauge connection is used to eliminate the time component
of the gauge connection from the stress tensor density in the static field
configuration. The energy, which is the integral of this static stress
tensor density, is bounded from below by a topological charge and is finite.
The static C-S solitons then are the solutions of the static second order
equations of this two dimensional subsystem.
The problem of gauging sigma models, which was considered sometime ago by 
Fadde'ev \cite{Faddeev} both in $2+1$ and in $3+1$ dimensions, is a 
pertinent one both physically and for its own sake. 
Recently there has been renewed interest in this area, 
thus far in $2+1$ dimensions. First the $C P^1$ model was gauged with a 
Chern-Simons $U(1)$ field in Refs.\cite{Aitchison}, and soon after that 
the $U(1)$ gauging of the $C P^1$ model with a Maxwell field, 
and including a Skyrme term was presented in Ref.\cite{Tchrakian}.

More recently Schroers\cite{Schroers} gauged the $O(3)$ sigma 
model with a Maxwell $U(1)$ field, much in the same spirit as in the 
earlier work of Ref.\cite{Faddeev}. The O(3) sigma model was gauged with
Chern-Simons dynamics in \cite{Arthur2}. It is interesting to ask if the
model supports attractive and repulsive phases. 

\section{The $O(3)$ Model}

We start by defining the Lagrangian on $2+1$ dimensional Minkowski space,
\be
{\cal L} = 
\frac{\kappa}{2\sqrt{2}} \varepsilon^{\mu \nu \rho} F_{\mu \nu} A_{\rho} +
(D_{\mu} \phi^a)^2 - 4V(\phi^3)
\label{e1}
\ee
where the three 
component field $\phi^a =(\phi^{\alpha} , \phi^3)$ with $\alpha =1,2,$ is 
constrained by $\phi^a \phi^a =1$, and the covariant derivative $D_{\mu} 
\phi^a$ is defined as in \cite{Tchrakian}, and with the opposite sign of the 
coupling used in \cite{Schroers}, as
\begin{equation}
\label{e2}
D_{\mu} \phi^{\alpha} =\partial_{\mu} \phi^{\alpha} + A_{\mu} 
\varepsilon^{\alpha \beta} \phi^{\beta} ,\qquad \qquad D_{\mu} \phi^3 
=\partial_{\mu} \phi^3 .
\end{equation}
The Lagrangian \rfa{e1} is $U(1)$ gauge invariant by virtue of \r{2} .
The potential function, not yet specified, is allowed only to depend 
only on the $U(1)$ invariant component $\phi^3$ of $\phi^a$. The energy
momentum tensor of this Lagrangian is
\be
\label{e3}
T_{\mu \nu} = 2 D_\mu \phi^a D_\nu \phi^a -
g_{\mu \nu} ( D_\mu \phi^a D_\nu \phi^a - 4 V)
\ee
The Hamiltonian density is given by;
\be
T_{00}= {\cal H} = D_0 \phi^a D_0 \phi^a + D_i \phi^a D_i \phi^a +
4 V + ( A_0 \phi^a )^2 .
\label{e4}
\ee

The essential feature of all Chern-Simons solitons of all $2+1$ dimensional
theories is that the time component of the Euler-Lagrange equation arising
from the variation of the gauge connection is used to eliminate the time
component of the gauge connection from the stress tensor density in the
static field configuration. In the static limit this is solved for $A_0$
yielding;
\be
A_0 = - \frac{\kappa}{2 \sqrt{2}} 
\frac{ \varepsilon^{ij} F_{ij} }{ (\phi^\alpha)^2 }.
\ee
In the static limit $T_{00} = {\cal H}$ reduces to;
\be
{\cal H} \left[ \frac{\kappa^2}{2} \frac{ F_{ij}^{2} }{ (\phi^\alpha)^2 } +
( D_i \phi^a )^2 + 4 V \right]
\label{ch4eq3}
\ee
The choice of potential function will be dictated by the requirement that the
volume integral of ${\cal H}$ is bounded below by a topological charge. This
fixes the potential uniquely. The topological charge density is not a total
divergence, but is only locally a total divergence. The usual winding number
density is 
$\varrho_0 =
\varepsilon_{ij} \varepsilon^{abc} \partial_i \phi^a \partial_j 
\phi^b \phi^c $, which is related to its gauged version 
\[
\varrho_1 =
\varepsilon_{ij} \varepsilon^{abc} D_i \phi^a D_j \phi^b \phi^c.
\]

The lower bound on the volume integral of the static Hamiltonian density
(\ref{ch4eq3}) can be inferred from the following inequalities;
\be
(\varepsilon_{ij} D_i\phi^a -\varepsilon^{abc} D_{j}\phi^b \phi^c )^2 
 \ge  0 , \:\:\:
(\frac{\kappa}{\sqrt{2} | \phi^\alpha |} F_{ij} -
\sqrt{2} \varepsilon_{ij} V )^2  \ge 0 .
\label{ch4eq6}
\ee
When the expansion of the static Hamiltonian density is examined, the
potential is specified uniquely by the requirement that the energy density
be bounded below by a total divergence. The lower bound 
can be arranged to be equal to the winding number density $\varrho_0$ 
plus a total divergence by identifying the potential uniquely as;
\[
V = \frac{1}{4 \kappa^2} (1 - \phi^3 )^3 (1 + \phi^3)
\]
The requirement that the surface integral should vanish is satisfied by
choosing appropriate boundary conditions. In the case of topologically 
stable solutions, the asymptotic conditions
\begin{equation}
\label{ch4eq8}
\lim_{|\vec x| \rightarrow 0}\phi^3 = -1,\:\:\:\:\:\:\
\lim_{|\vec x| \rightarrow \infty}\phi^3 =  \pm 1
\end{equation}
guarantee that the volume integral of $\varrho$ yields a nonzero 
integer winding number. This statement assumes that $A_i$ does not grow 
too fast at infinity, an assumption which is amply justified as will be seen 
below when we specialise to the radial field configuration. The conditions 
\rfa{ch4eq8} with the {\it upper sign} pertain to the {\it topologically stable} 
solutions of nonzero winding, while those with the {\it lower sign} 
to the {\it nontopological} vortices. In the latter case, the winding 
number vanishes for all vorticities $N$,
in this case the energy is given a lower bound by the magnetic flux.

\section{Self Dual Solutions}

The topological inequality is saturated when the inequalities
\rfa{ch4eq6} are saturated (for the sake of definiteness we have
chosen the value of $\kappa=1$ ), yielding the Bogomol'nyi equations,
\begin{equation}
\label{ch4eq9}
F_{ij} = \mp \varepsilon_{ij} (1 - \phi^3)^2 (1+\phi^3)
\end{equation}
\begin{equation}
\label{ch4eq10}
\varepsilon_{ij} D_i \phi^a = \pm \varepsilon^{abc} D_j \phi^b \phi^c ,
\end{equation}
where the lower/upper signs pertain to anti/self-duality. Our radially 
symmetric Ansatz for the fields $A_i$ and $\phi^a$ is
\begin{equation}
\label{ch4eq11}
A_i =\frac{a(r)-N}{r} \varepsilon_{ij} \hat x_j
\end{equation}
\begin{equation}
\label{ch4eq12}
\phi^{\alpha} =\sin f(r) \: n^{\alpha} , \: \phi^3 =\cos f(r)
\end{equation}
where $\hat x_i =\frac{x_i}{r}$ and $n^{\alpha} =(\cos N\theta ,\sin 
N\theta)$ are unit vectors, with $N$ defined to be an integer.

The Bogomol'nyi equations \rfa{ch4eq9} and \rfa{ch4eq10} now reduce to the following 
pair of coupled nonlinear first order ordinary differential equations;
\begin{equation}
\label{ch4eq13}
\frac{a'}{r}= \pm (1-\cos f)^2 (1-\cos f),
\qquad \qquad f'=\pm \frac{a \sin f}{r}.
\end{equation}
The {\it topological} asymptotic conditions, namely \rfa{ch4eq8} with the upper 
sign, which were chosen in 
anticipation of our restriction to the antiselfdual case, now read
\begin{equation}
\label{ch4eq14}
\lim_{r \rightarrow 0}f(r) = \pi,\:\:\:\:\:\:\lim_{r \rightarrow 
\infty}f(r) = 0
\end{equation}
which for the field configuration \rfa{ch4eq11} and  \rfa{ch4eq12}
imply {\it vorticity} $=-N$. This is 
the same as in the usual (ungauged) $O(3)$ model where the radially 
symmetric antiselfdual vortices satisfy the asymptotic conditions
\rfa{ch4eq14} , while the selfdual vortices satisfy instead 
$\lim_{r \rightarrow 0}f(r) = 0;\:\: \lim_{r \rightarrow \infty}f(r) =\pi $. 

The asymptotic behaviour of the function $a(r)$ in \rfa{ch4eq11} is of no 
consequence to the topological stability of the soliton, unlike in the 
cases of the Higgs models \cite{Hong}, and of 
the gauged $\C P^1$ models \cite{Aitchison} \cite{Tchrakian}. This is because in 
the latter systems the topological charge, which is again related to the 
vorticity, is also proportional to the {\it magnetic flux}. In the case of 
the gauged $O(3)$ models, defined in \cite{Schroers} and here, the magnetic 
flux of the solution is not restricted by the requirement of the stability of 
the soliton. The only constraint on the large $r$ behaviour of $a(r)$ here 
is the requirement that the surface integral bounding below the static
Hamiltonian density, should 
vanish. This means that $a(r)$ should not grow faster than the quantity 
$(\cos f -1)$ in that region. Since the magnetic flux is proportional to the 
quantity$[-a(\infty) +a(0)]$,
we shall seek solutions for which both $a(\infty)$ and $a(0)$ are finite, 
since it is reasonable that the solutions we seek correspond to finite 
magnetic flux field configurations. As explained above, we shall take 
$a(0)=N$, but will take $a(\infty)=\alpha$, where $\alpha$ is a nonzero 
constant whose sign will depend on whether we are considering the 
topological or the nontopological solutions. 
Corresponding to the asymptotic conditions \rfa{ch4eq14} for the function $f(r)$, 
we state the asymptotic conditions on the function $a(r)$ as
\begin{equation}
\label{ch4eq15}
\lim_{r \rightarrow 0}a(r) = N,\:\:\:\:\:\:\lim_{r \rightarrow 
\infty}a(r) = \alpha .
\end{equation}

\section{Non-Self dual solutions}

We will not study further the Bogomol'nyi equations, as they have been amply
remarked upon in Refs. \cite{Ghosh} \cite{Kimm} \cite{Arthur2}.
It should be noted that in the latter reference, \cite{Arthur2},
the proof of existence of the solitons has been demonstrated. 
The $N=1$ soliton exists, in contrast to the proof of nonexistence of the
$N=1$ soliton which has been given in \cite{Schroers} in the case of the
O(3) sigma model with a Maxwell term describing the curvature.

We will examine the Euler-Lagrange equations arising from the 
static Hamiltonian. Using the Ansatz the static Hamiltonian reduces to the
one dimensional subsystem:
\be
L = r \left[ \left( \frac{a_r}{r \sin f} \right)^2 +
f_r^2 + \left( \frac{ a \: \sin f}{r} \right)^2 +
\lm_0 (\sin f(1 - \cos f ))^2 
\right],
\label{ch4eq16}
\ee
defined by 
\be
\int d^2 x \: {\cal H} = 2 \pi \int dr \: L.
\label{ch4eq17}
\ee
When the one dimensional Lagrangian is varied the resulting equations are;
\bean
\label{ch4eq18}
a(r) \sin^4 f + 2 \cot (f) a_r f_r + \frac{a_r}{r} - a_{rr} & = & 0 \\
\lm_0 r(1-\cos f) \sin f (\cos f - \cos^2 f + \sin^2 f ) +
\frac{a^2 \cos f \sin f}{r} &-& \\
 \frac{\cot f \csc^2 f a_{r}^{2}}{r} - f_r - r f_{rr} &=& 0,
\ean
where $f_r = \frac{df}{dr}$. We first examine the equations \rfa{ch4eq18} in
the region $r \ll 1$ region. The solution is;
\bea
f(r) &=& \pi + f_o r^N + f_1 r^{N+2} + O(r^{N+4}) \\
a(r) &=& N + A_o r^{2N+2} + A_1 r^{2N+4} + O(r^{2N+6}) \mbox{ where,} \\
\mbox{If } N & = & 1 \left\{ \begin{array}{l}
f_1 = - \frac{ 24 A_o^2 + f_o^6 - 6 f_o^4 \lm_0 }{12 f_o^3}, \\ 
A_1 = - \frac{ 8 A_o f_o^3 - 3 f_o^5 - 48 A_o f_1 }{36 f_o}, 
\end{array} \right. \\
\mbox{If } N & \ge & 2 \left\{ \begin{array}{l}
f_1 = - \frac{A_o^2 (1+N)^2 - f_o^4 \lm_0 }{ (1+N) f_o^3}, \\ 
A_1 =   \frac{ 2 A_o f_1 (1+N)}{(2+N) f_o}, 
\end{array} \right.
\eea
The constants $f_o$ and $A_o$ are fixed by the asymptotic value of the 
solution at infinity. They are found by using numerical methods, which
require correct values of the fields at the boundary.

Using a power series solution of the Bogomol'nyi equations, it is found that
the ratio
\be
\frac{A_o}{f_o^2} = \frac{1}{N+1}
\label{ch4eq18a}
\ee
holds. This gives one free parameter in the analytic series which controls 
the behaviour of the solutions. It will also serve as a check on the 
numerically evaluated constants.

Now considering the region $r \gg 1$ and anticpating decaying solutions,
we linearise the Euler Lagrange equations about their asymptotic values. That
is, the equations are linearised in the functions $F(r)$ about $f=0$ and
$A(r)$ around the asymptotic value $a(r) = \alpha $. 
The Euler Lagrange equation for $f(r)$ is linearised about its asymptotic
value of zero, and found to be;
\[
\alpha^2 F - r F_r - r^2 F_{rr} = 0.
\]
The solutions to this are ;
\[
F = \frac{c_1}{r^\alpha}+ c_2 r^\alpha.
\]
In order to have finite energy solutions the constant $c_2$ is chosen as
zero. The equation for $a(r)$ is also linearised, and found to be;
\[
a_{rr} - r a_r = 0,
\]
whose solution is $a= c_3 r^2$ or that $a$ is a constant. The latter option
is chosen, in order to restrict attention to finite energy solutions. This
constant is chosen to be $\alpha$. In the
case of the analysis of the Bogomol'nyi equations there is found to be a
restriction of the value of the constant $\alpha$. It satisfies the inequality;
$\alpha > 1$.

\section{Numerical Solutions}

We have studied the equations numerically using a shooting method. The
integration started in the region $r \ll 1$ using as initial data the power
series solutions. The constants $f_o$ and $A_o$ have been found which give
the correct behaviour, as $r \gg 1$, of the functions $f(r)$ and $a(r)$.
The profiles for the functions $f(r)$ for
vorticity $N=2,4$ and those for $a(r)$ are given in figures 1,2,4,5, and
the respective energy density profiles in figures 3 and 6. In table 3 the
total energies of the vortices have been calculated for the approximate
solutions. If the Bogomolny'i equations are substituted into the one
dimensional system the result is;
\be
L_0 = \frac{2 f_r a \sin f}{r} + \frac{2 a_r}{r} (1- \cos f) 
\ee
The total energy of a topological self dual solution can be written as;
\bea
E_{sd} &=&
4 \pi \int_{0}^{\infty} \: dr \: \frac{d \:}{dr} \left( a(1-\cos f) \right) \\
&=& 8 \pi | N | = 2 \pi ( 4 | N | ).
\eea
This last line is found using the boundary conditions for both functions
$a(r)$ and $f(r)$. The energy is independent of the choice of the number
$\alpha$. It is simply enough that $a_r (r \ra \infty) = 0$.
The figures for the energy density, in both graphs and tables, are given
in units of $ 2 \pi$. The figures of the energy density for the value of
constant $\lm_0 = 1$ is the self dual limit, $4 |N|$ in our convention. It
can be seen that the calculated values are close approximations
to the analytic calculation. In figure 7, the values of total energy
are plotted. For the $n=2$ vortex, these figures are multiplied by 2. It can 
be seen that the $n=4$ vortex energy is larger than the $n=2$ energy,
for the regime where $\lm_0 < 1$. 
Also it can be seen that the $n=2$ vortex energy is larger than the
$n=4$ energy, for the regime where $\lm_0 > 1$. There is a, "cross over",
between attraction and repulsion behaviour. The relations;
\[
2 {\cal E}(n=2,\lm_0 <1) < {\cal E}(n=4,\lm_0 <1), \: \: \:
2 {\cal E}(n=2,\lm_0 >1) > {\cal E}(n=4,\lm_0 >1),
\]
can be seen from the data, and in the graph. This states that a vortex with
degree 2 is lighter that a vortex with degree 4, so it is energetically
favourable that vortices repel. Clearly in the second relation, attraction
is favourable. The boundary conditions for the vortices were chosen to be
the anti-selfdual configuration. When the vortices are self dual, that is
when $\lm_0 =1$ the energy of 4 vortices is twice that of 2 vortices.
Only at the value of the constant
$\lm_0 =1 $ are the energies per unit vortex number equal.
This means that there is no interaction energy
between self dual vortices. It can be shown that the diagonal components of
the stress tensor vanish in the self dual configuration. The values
of the constants $A_o$ and $f_o$ for the self dual solution should be related
by equation \rfa{ch4eq18a}. It is seen that the numerically calculated
numbers are close to their theoretical ratio.

For the nontopological solution, different boundary conditions are used and
the result of calculating the energy is $E_{sd} = 4n + 4 \alpha$. Although
the value of the energy in the nontopological case depends on the surface
integral contribution and is, in fact, the magnetic flux, the energy is
arbitrary. We have not examined the nontopological solitons, because of this.

When the present work was completed, the work of Ref. \cite{Gladikowski} was
brought to my attention. Our approaches to the problem of the gauged O(3) 
sigma model are widely different, in particular, in Ref. \cite{Gladikowski},
the examination of varying a different a coupling constant is made.
The present work is different, in the sense that it was carried out in the
spirit of Ref. \cite{Jacobs}, where  the Higgs coupling constant is varied.

{\large Acknowledgement:} 

It is a pleasure to thank Prof. D.H. Tchrakian,
who suggested the problem, and for continued helpful discussions.

\newpage
\begin{tabular}{|l|l|l|} \hline \hline
{\em Constants} & \multicolumn{2}{c|}{$n=2$} \\ \hline
$\lambda$  &  $ A_o $ & $ F_o $  \\ \hline
0.8 & -7.90372 & -5.0000 \\
0.9 & -8.09182 & -5.0000 \\ 
1.0 & -8.33333 & -5.0000 \\
1.1 & -8.53843 & -5.0000 \\
1.2 & -8.65862 & -5.0000 \\ \hline \hline
\end{tabular}

\vspace{2cm}

\begin{tabular}{|l|l|l|} \hline \hline
{\em Constants} & \multicolumn{2}{c|}{$n=4$} \\ \hline
$\lambda$  &  $ A_o $ & $ F_o $  \\ \hline
0.8 & -4.68536 & -5.00000 \\
0.9 & -4.85028 & -5.00000 \\
1.0 & -5.00000 & -5.00000 \\
1.1 & -5.13790 & -5.00000 \\
1.2 & -5.26629 & -5.00000 \\ \hline \hline
\end{tabular}

\vspace{2cm}

\begin{tabular}{|l|l|l|} \hline \hline
{\em Constants} & \multicolumn{2}{c|}{Energy} \\ \hline
$\lambda$  &  $E(n=2,\lambda)$ & $E(n=4,\lambda)$  \\ \hline
0.8 & 7.9408 & 15.9423 \\
0.9 & 7.9694 & 15.9695 \\
1.0 & 8.0000 & 16.0000 \\
1.1 & 8.0411 & 16.0330 \\
1.2 & 8.0819 & 16.0680 \\ \hline \hline
\end{tabular}

\newpage

\noindent Figure 1:
Profile of the function $f(r)$ for the vortices with $n=2$ with
$\lambda_{0}=1.2,\ldots,0.8$.

\noindent Figure 2:
Profile of the function $a(r)$ for the vortices with $n=2$ with
lower values of $\alpha$ corresponding to lower $\lm_0$.

\noindent Figure 3:
Profile of the energy density for the $n=2$ vortices where increasing
peaks represent increasing energy and increasing $\lambda_{0}$.

\noindent Figure 4:
Profile of the function $f(r)$ for the vortices with $n=4$ with
$\lambda_{0}=1.2,\ldots,0.8$.

\noindent Figure 5:
Profile of the function $a(r)$ for the vortices with $n=4$ with
lower values of $\alpha$ corresponding to lower $\lambda_{0}$.

\noindent Figure 6:
Profile of the energy density for the $n=4$ vortices where increasing
peaks represent increasing energy and increasing $\lambda_{0}$.

\noindent Figure 7:
Graph of the energy of two superimposed vortices,
$2 \times {\cal E}(\lambda,n=2)$ and ${\cal E}(\lambda,n=4)$,
as a function of $\lambda_{0}$. 

\end{document}